\documentclass{aa}  

\usepackage{graphicx}
\usepackage{txfonts}
\usepackage{soul}
\usepackage{hyperref}
\hypersetup{
    colorlinks   = true, %Colours links instead of ugly boxes
    urlcolor     = cyan, %Colour for external hyperlinks
    linkcolor    = blue, %Colour of internal links
    citecolor   = blue %Colour of citations
}
\usepackage{threeparttable}
\usepackage{placeins}
\usepackage{xcolor}
\usepackage{soul} % for strikethrough text
\setstcolor{red} % striking in red
\graphicspath{plots}

\begin{document}
\title{\emph{Young, wild and free}: the early expansion of star clusters}
\subtitle{}

\author{
    A. Della Croce \inst{1,2}\thanks{\email{alessandro.dellacroce@inaf.it}}
    \and
    E. Dalessandro\inst{2}
    \and 
    A. Livernois\inst{3}
    \and 
    E. Vesperini\inst{3}
}

\institute{
    Department of Physics and Astronomy ‘Augusto Righi’, University of Bologna, via Gobetti 93/2, I-40129 Bologna, Italy
    \and
    INAF – Astrophysics and Space Science Observatory of Bologna, via Gobetti 93/3, I-40129 Bologna, Italy
    \and 
    Department of Astronomy, Indiana University, Swain West, 727 E. 3rd Street, IN 47405 Bloomington, USA
}

\date{Received \dots; accepted \dots}
 
\abstract
{
Early expansion plays a fundamental role in the dynamical evolution of young star clusters.
However, until very recently most of our understanding of cluster expansion was based only on indirect evidence or on statistically limited samples of clusters.
Here we present a comprehensive kinematic analysis  
of virtually all known young ($t<300$ Myr) Galactic clusters based on
the improved astrometric quality of the Gaia DR3 data.
Such a large sample provides the unprecedented opportunity to robustly constrain the fraction of clusters and the timescale during which expansion has a prominent impact on the overall kinematics. 
We find that a remarkable fraction (up to $80\%$) 
of clusters younger than $\sim30$ Myr is currently experiencing significant expansion, whereas older systems are mostly compatible with equilibrium configurations. 
We observe a trend where the expansion speed increases with the clustercentric distance, suggesting that clusters undergoing expansion will likely lose a fraction of their present-day mass. 
Also, most young expanding clusters show large sizes, possibly due to the expansion itself.
% Also, we speculate that expanding clusters have larger sizes as the result of the expansion itself. 
% Also, we find that expanding clusters have typically large sizes. 
A comparison with a set of N-body simulations of young star clusters shows that the observed expansion pattern is in general qualitative agreement with that found for systems undergoing violent relaxation and evolving toward a final virial equilibrium state. However, we also note that additional processes likely associated with residual gas expulsion and mass loss due to stellar evolution are also likely to play a key role in driving the observed expansion.
% Also, we observe a positive trend of the expansion speed with the clustercentric distance, suggesting that clusters undergoing expansion will likely lose a fraction of their present-day mass. Finally, as expected, we find that expanding clusters show larger sizes. 
% A comparison with a set of $N$-body simulations of young star clusters suggests that the observed expansion pattern of a \textbf{significant fraction of clusters} is qualitatively in agreement with the expectations for systems undergoing violent relaxation and evolving toward a final virial equilibrium state.  
% Additional processes likely associated with primordial gas expulsion and mass loss due to stellar evolution are likely to play a role in driving the observed expansion.
}

\maketitle
%
%-------------------------------------------------------------------

\section{Introduction}
It is commonly accepted that most Galactic star formation takes place in embedded proto-clusters as a result of turbulent, clumpy, and stochastic processes in molecular clouds 
\citep[e.g.][]{lada&lada_2003,offner_etal2009,feigelson_etal2013}. 
The 
environments in which stars form determine to some extent 
a number of key properties characterizing the stellar clusters we observe today, such as their initial mass function, the stellar multiplicity, and the star-by-star chemical abundance differences routinely observed in massive stellar clusters
\citep[e.g.][]{bastian&lardo_2018,gratton_etal2019,milone&marino2022}.

The past years have witnessed a renewed interest in the theoretical and observational study of the formation of stellar clusters.
Particular attention has been paid to the very early evolutionary phases,
when clusters are expected to undergo violent relaxation and
to be evolving toward a virial equilibrium state.
Studies of clusters’ formation and very early evolution have shown that clusters can emerge from these evolutionary phases with 
internal rotation, radial anisotropy in the velocity distribution, and mass segregation.
However, the detailed role of the various physical processes involved and the variety of different dynamical paths are still a matter of intense investigation 
\citep[see e.g.][]{mcmillan_etal2007,allison_etal2009,moeckel&bonnell_2009a,pfalzner_Kaczmarek_2013,
Banerjee&kroupa_2014,vesperini_etal2014,fujii&portegiesZwart_2016,parker_etal2016,dominguez_etal2017,mapelli_2017,daffner-powell&parker2020,gonzalez-samaniego&Vazquez-Semadeni_2020}. 

The clusters' early evolutionary phases have been poorly constrained also by observations, mainly due to the difficulty of obtaining reliable kinematic information for statistically significant samples of cluster members.
Data from the Gaia space mission and, in particular the DR3 \citep{gaiadr3}, provided the deepest and most precise astrometric catalog ever obtained and opened the possibility of new studies about star cluster formation and dynamics.
Indeed, Gaia's unprecedented kinematical mapping of the Galaxy 
and of its stellar components significantly enriched our knowledge about the number and the properties of star clusters in the Milky Way 
\citep[see e.g.][]{cantat-gaudin+20,cantat-gaudin&anders_2020}. 
Also, it enabled detailed studies of nearby star-forming regions and their use as ideal laboratories to shed new light on cluster formation and early evolution 
\citep[e.g.][]{beccari_etal2018,kuhn_etal2019,meingast_etal2019,dalessandro_etal2021,dellacroce_etal2023}.

The evolution of a young cluster depends on many physical factors, such as the cluster’s structural properties (e.g. mass and size), the strength of the external tidal fields, and their physical 
and dynamical ages \citep[see e.g.][]{hills1980,parker_etal2014,sills_etal2018}.
Early cluster expansion is expected to play a fundamental role
in cluster evolution and to drive stellar systems toward complete dissolution
\citep{elmegreen_1983,mathieu_1983,adams_2000,kroupa2001_imf,goodwin_etal2006,Pelupessy_portegiesZwart2012,dinnbier_etal2020,dinnibier_etal2022,farias_etal2023}. 
Theoretical studies show that cluster expansion due to processes like gas expulsion and the subsequent violent relaxation phase may last several Myr, depending on many cluster properties such as the cluster dynamical state at the expulsion, the specific star-formation efficiency, and cluster size, density, and mass
\citep{kroupa_etal2001,baumgardt_kroupa2007,Pelupessy_portegiesZwart2012,banerjee&kroupa2013,pfalzner_Kaczmarek_2013,brinkmann_etal2017,farias_etal2017,farias_etal2023,li_etal2019,pang2021,leveque_etal2022}.
Eventually, the system can either settle into equilibrium, after losing some mass during expansion or dissolve completely. 
The subsequent long-term evolution of surviving clusters will be driven by two-body relaxation and the interplay between the effects of internal evolution and the external Galactic tidal field \citep[see e.g.][]{deLaFuenteMarcos1997,Hurley_etal2005}.

However, evidence of cluster expansion has been hard to derive directly, and until very recently most of our understanding 
has been based on indirect evidence resulting from cluster size$-$density or density$-$age anti-correlations \citep[e.g.][]{pfalzner_etal2014,getman_etal2018}. 
Thanks to Gaia data we are now able to directly probe expansion in star-forming regions and young star clusters with unprecedented detail \citep{CantatGaudin_etal2019_jordi,CantatGaudin_etal19,romanZuniga_etal2019,damiani_etal2019,Wright_etal2019,lim_hong_etal2020,lim_naze_hong_etal2021,lim_naze_hong_etal2022,buckner_etal2020,armstrong_etal2020,armstrong_etal2022,schoettler_etal2020,Schoettler_etal2022,kuhn_etal2020,swiggum_etal2021,mainz-appellaniz2022,miret-roig_etal2022,Guilherme-Garcia_etal2023}.

In particular, \citet{kuhn_etal2019} 
studied the kinematical properties of a sample of 28 young ($1-5$ Myr) clusters and associations by using Gaia (DR2) proper motions.
According to those authors, 
observations are consistent with early cluster expansion driven by changes in the gravitational potential due to the dispersal of the molecular cloud.
More recently, \citet{Guilherme-Garcia_etal2023} investigated the kinematics of 1237 clusters using a technique that aims at reconstructing the underlying velocity field. They found 8 (and an additional 9 candidates) clusters displaying rotation patterns whereas 14 (15 more candidates) clusters show evidence of expansion in their velocity fields \citep{Guilherme-Garcia_etal2023}. The vast majority of the expanding systems in their sample are younger than 100~Myr, with a peak around 10~Myr.
While these studies provided important clues about the early evolution and survival of young clusters, the availability of more precise data from the Gaia DR3 and the possibility to assess cluster membership more robustly can allow us to disentangle critical aspects related to their evolution.

In this work, we present a comprehensive and systematic analysis of the kinematical properties of (virtually) all young ($t<300$ Myr) stellar clusters identified so far in the Milky Way \citep[using the cluster catalog by][]{cantat-gaudin&anders_2020,cantat-gaudin+20}, with particular focus on the early cluster expansion phase.
Such a large sample provides the unprecedented opportunity to robustly constrain the timescale during which expansion has a prominent impact on the overall cluster kinematics. 
Also, it allows us to trace how expansion affects or is linked to the cluster properties and formation mechanisms.

The paper is organized as follows. In Section~\ref{sec:data_analysis} the cluster membership determinations and kinematic analyses are described. We present our results in Section~\ref{sec:results} and we compare them with numerical simulations in Section~\ref{sec:numerical_simulations}. We assess the impact of different age estimates on our results in Section~\ref{appendix:different_age_estimate} and a detailed comparison with previous works is presented in Section~\ref{sec:comparison_previous_works}. Finally, conclusions are drawn in Section~\ref{sec:conclusions}.

\section{Data Analysis}\label{sec:data_analysis}
\subsection{Clusters' membership}\label{sec:membership}
\begin{figure*}[!ht]
    \centering
    \includegraphics[width=\textwidth]{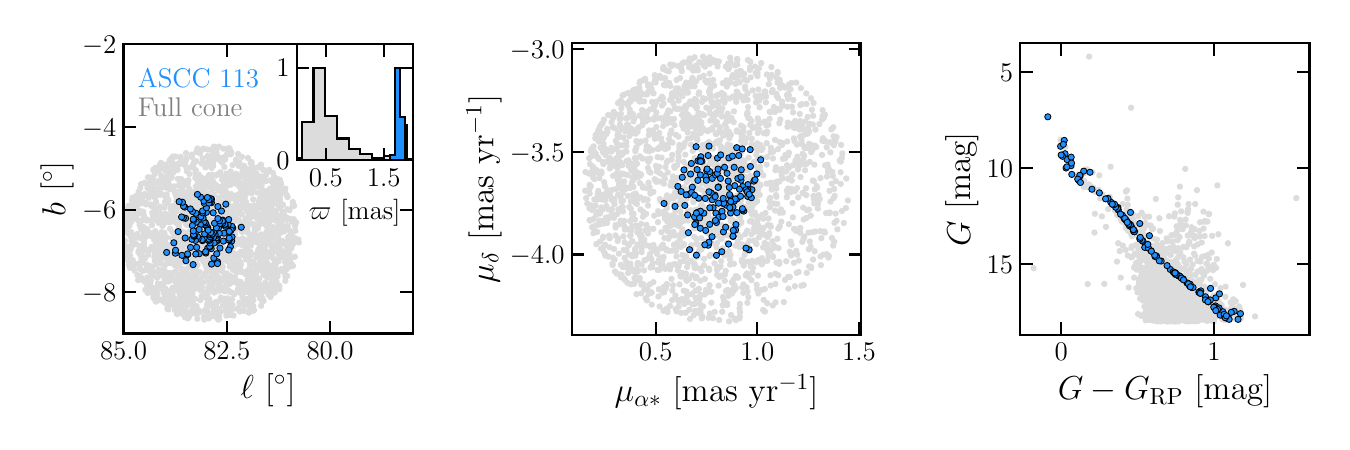}
    \caption{
    Astrometric and photometric properties of ASCC~113 members (shown in blue), whereas in gray we show the properties of the initial sample of Gaia sources (see Section~\ref{sec:membership}). 
    In the left panel are the distributions in Galactic coordinates and parallax (top-right corner), while the middle panel shows the PM distribution. Finally, the color-magnitude diagram is shown in the right panel. The parallax distributions were scaled for visualization purposes only.
    }
    \label{fig:diagnostic_ASCC113}
\end{figure*}
The present work makes use of the list of clusters identified by \citet{cantat-gaudin+20} using Gaia DR2 data. We focused on systems younger than $\leq 300$ Myr \citep[according to age estimates by][]{cantat-gaudin+20} to investigate the evolution of star clusters 
in their very early stages. In this way, we selected 1179 clusters out of 2017 in the original catalog.

To take full advantage of the most recent and accurate Gaia DR3 data release for the definition of cluster member stars, we performed an independent membership analysis.
For every cluster, we retrieved Gaia DR3 data for sources brighter than $G=18$, and that have a 5-parameter solution (i.e. with sky position, proper motions, and parallax measurements). In particular, every query was centered on the cluster's centroid (as reported by \citealt{cantat-gaudin+20}) 
and the search radius was defined as $R_{\rm search,\,sky} \equiv 2 R_{95,\,{\rm sky}}$, where $R_{95,\,{\rm sky}}$ 
is the radius enclosing 95\% of the member stars reported by \citet{cantat-gaudin+20}.

For each cluster, we then selected stars according to their motion with respect to the cluster's bulk motion. 
Only sources within $R_{\rm search,\,PM} \equiv 2\,R_{95,\,{\rm PM}}$, with $R_{95,\,{\rm PM}}$ being the circle in proper motion space enclosing 95\% of 
the members were retained in the subsequent analysis.
We did not apply any preliminary parallax selection.
These selections allowed us to include all the previously listed members in our starting Gaia DR3 catalog. 

We performed the clustering analysis in the five-dimensional space of Galactic coordinates, proper motions, and parallax 
(i.e. $\ell$, $b$, $\mu_{\rm \alpha *}$, $\mu_\delta$, and $\varpi$). 
Since we are dealing with heterogeneous quantities, we preliminary scaled them all,
so that the mean and standard deviation of their distributions are equal to zero and one respectively.
We used the \texttt{StandardScaler} provided by the Python library \texttt{sklearn}.
We then performed an unsupervised clustering analysis on the scaled coordinates by means of the \texttt{HDBSCAN} 
\citep{hdbscan,hdbscan-accelerated} algorithm.

Based on the results of several tests, we set the algorithm's parameters as follows.
\begin{align}
    \texttt{min\_cluster\_size} = 
    \begin{cases}
        20  & \text{for $N_{\rm CG20}<100$}\,,\\
        \frac{N_{\rm CG20}-100}{10}+50 & \text{for $N_{\rm CG20}\geq100$}\,,
    \end{cases}
    \label{eq:minClusterSize_hdbscan}
\end{align}
and
\begin{align}
    \texttt{min\_samples} = 
    \begin{cases}
        5  & \text{for $N_{\rm CG20}<100$}\,,\\
        10 & \text{for $N_{\rm CG20}\geq100$}\,,
    \end{cases}
    \label{eq:minSamples_hdbscan}
\end{align}
where $N_{\rm CG20}$ is the number of member stars reported by \citet{cantat-gaudin+20}.
These parameters appeared to be best suited for the unsupervised search for members in star clusters 
characterized by significantly different extensions ($R_{\rm search,\, sky}$), velocity distributions ($R_{\rm search,\, PM}$) and number 
of likely members ($N_{\rm CG20}$). 

As a test case, in Fig.~\ref{fig:diagnostic_ASCC113} we show the spatial, parallax, proper motion, and color-magnitude diagram (CMD) 
distributions for the selected members of ASCC~113 
% (whose kinematic properties are presented in Fig.~\ref{fig:streamplot_ASCC113}). 
As expected, selected member stars are clustered in the five-dimensional astrometric space 
(i.e. $\ell;b;\varpi;\mu_{\alpha *};\mu_\delta$).
Also, they exhibit a well-defined, cluster-like sequence in the CMD,
thereby confirming the ability of our clustering analysis to recover stellar cluster members among field stars.

\begin{figure}[!ht]
    \centering
    \includegraphics[width=0.4\textwidth]{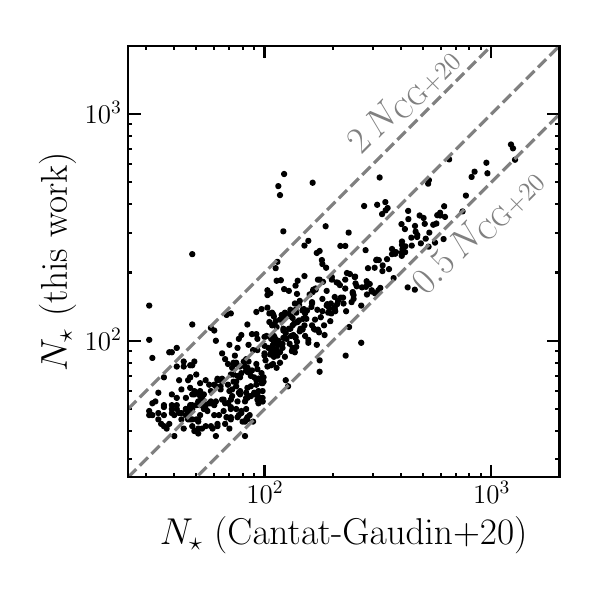}
    \caption{Comparison of member stars between this work and \citet[][$N_{\rm GC+20}$]{cantat-gaudin+20}, for all the clusters included in the subsequent analysis. Diagonal lines show the one-to-one relation and the relations corresponding to half and twice the members (as also reported in the plot).
    }
    \label{fig:member_comparison}
\end{figure}

While running the procedure for all the clusters in the sample, we found that the algorithm performed poorly 
on clusters with only a few member stars ($N_{\rm CG20}<30$). 
This is probably due to the choice of \texttt{min\_cluster\_size}, which sets a lower limit to the number of member stars in our search for star clusters. 
However, since the kinematic analysis strongly benefits from the availability of relatively large samples of individual velocities, 
we decided to exclude these clusters from our analysis. 
In this way, we ended up with a final sample of 949 clusters for which the clustering analysis was performed.

Fig.~\ref{fig:member_comparison} shows a direct comparison between the number of member stars obtained by using Gaia DR3 and that found by \citet{cantat-gaudin+20}. 
While some differences are expected due to the different adopted clustering algorithms and Gaia data releases, Fig.~\ref{fig:member_comparison} shows an overall agreement between the two compilations.

\subsection{Cluster kinematics analysis}\label{sec:cluster_kin_prop}
% To investigate the evolution of star clusters in their early stages, we focused on the 1179 Galactic systems younger than $300$~Myr identified by \citet{cantat-gaudin+20}.
First, we accounted for perspective effects induced by the cluster bulk motion following the equations
reported by \citet{vanLeeuwen2009}. 
To this aim, systemic line-of-sight (LOS) velocities from \citet{tarricq_etal2021} were used. Only 509 out of 949 clusters (see Section~\ref{sec:membership}) had reported LOS velocity and were thus retained in the subsequent analysis. 
\begin{figure}[!ht]
    \centering
    \includegraphics[width=0.4\textwidth]{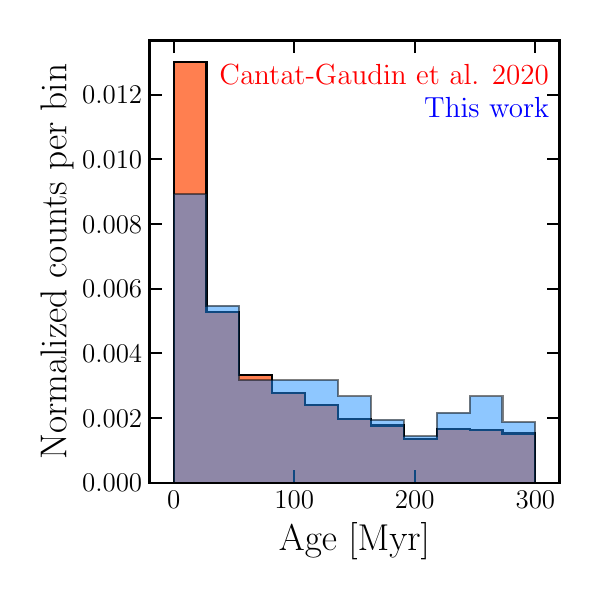}
    \caption{Age distributions for clusters in the starting catalog \citep[in red,][]{cantat-gaudin+20} and those retained in the kinematic analysis (in blue). Histograms were normalized such that their areas sum to unity.
    }
    \label{fig:agedistr_comparison}
\end{figure}
In Figure~\ref{fig:agedistr_comparison} we compare the age distributions within 300 Myr of the original catalog and the final sample of clusters. 
The distributions populate the investigated age range in a similar fashion and our final sample of 509 clusters is representative of the initial age distribution.

We then used stars with a membership probability larger than $70\%$ 
based on the clustering analysis performed in this work (Section~\ref{sec:membership}).
We also selected cluster members having \texttt{ruwe} $\leq1.4$, 
\texttt{astrometric\_gof\_al} $\leq1$, and \texttt{astrometric\_excess\_noise} $\leq1$ mas (if \texttt{astrometric\_excess\_noise\_sig} $>2$), thus excluding stars for which the standard five-parameter solution did not provide a reliable fit of the observed data \citep{Lindegren_etal2021}. 

We inferred the mean radial velocity, $\langle v_{\rm R}\rangle$, and the radial velocity dispersion, $\sigma_{\rm R}$, 
in a fully Bayesian framework properly accounting for errors on individual velocities. We explored the parameters space employing a Markov Chain Monte Carlo (MCMC) technique. In particular, we used the Python package \texttt{emcee} \citep{foreman-mackey_etal2013}.
For each system, we assumed the likelihood \citep{pryor_meylan_1993}
\begin{equation}
    \ln\mathcal{L} = -0.5\sum_{\rm k}\left[ \frac{(v_{\rm R,k} - \langle v_{\rm R}\rangle)^2}{\delta v^2_{\rm R,k} + \sigma^2_{\rm R}} + \ln(\delta v^2_{\rm R,k} + \sigma^2_{\rm R}) \right]\,,
    \label{eq:likelihood}
\end{equation}
with $v_{\rm R,k}$ and $\delta v_{\rm R,k}$ being the radial velocity and the respective error of the $k$-th member.
Eq.~\ref{eq:likelihood} assumes that the intrinsic distribution along the radial component of the velocity is a Gaussian with mean $\langle v_{\rm R}\rangle$ and velocity dispersion $\sigma_{\rm R}$.
We used uniform priors within $[-10;+10]$ mas yr$^{-1}$ and $[0.001;15]$ mas yr$^{-1}$ 
for $\langle v_{\rm R}\rangle$ and $\sigma_{\rm R}$ respectively.
For each cluster, we initialized 50 walkers and ran the algorithm for 500 steps, which was found to be sufficient 
for both ensuring convergence (for which the first half of samples was discarded) and accounting for correlations between samples (typically of the order of 20).
Median values and 16\% and 84\% quantiles (corresponding to the $1\sigma$ interval if the distributions were Gaussian) were then computed for each quantity directly from posterior samples.
The kinematical analysis was performed by adopting the clusters' geometric center, defined as the median of the positions of member stars. 

\section{Results}\label{sec:results}
\begin{figure*}
    \centering
    \includegraphics[width=\textwidth]{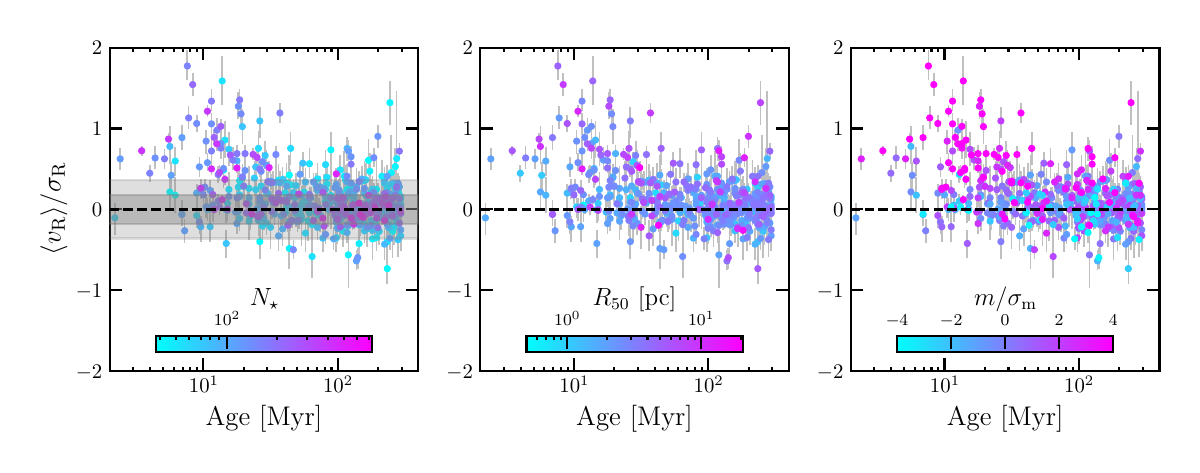}
    \caption{The ratio between the mean radial velocity and the radial velocity dispersion for clusters younger than 300 Myr \citep[according to][]{cantat-gaudin+20}. Errors on the $y$-axes were obtained directly from the MCMC samples, while ages are from \citet{cantat-gaudin+20}.
    In the left panel, clusters are color-coded according to the number of members. 
    Also, the standard deviation (and twice the value) of the $\langle v_{\rm R} \rangle / \sigma_{\rm R}$ ratio obtained from numerical realizations of equilibrium star clusters (see Section~\ref{appendix:montecarlo}) are shown as gray shaded areas. Values obtained from numerical realizations were convolved with the median observational error to allow for a direct comparison with the underlying data.
    In the middle panel colors depict $R_{\rm 50}$, whereas in the right panel, the color coding represents the ratio between the slope and the corresponding error from the linear regression of radial velocities as a function of cluster-centric distance.
    }
    \label{fig:expansion}
\end{figure*}
\begin{figure*}[!h]
    \centering
    % NGC 6193
    \includegraphics[width=.325\textwidth]{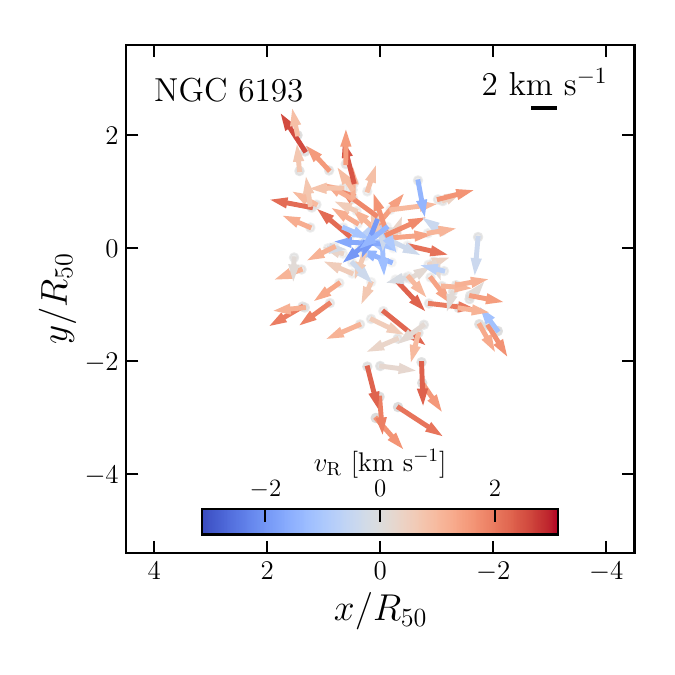}
    \includegraphics[width=.325\textwidth]{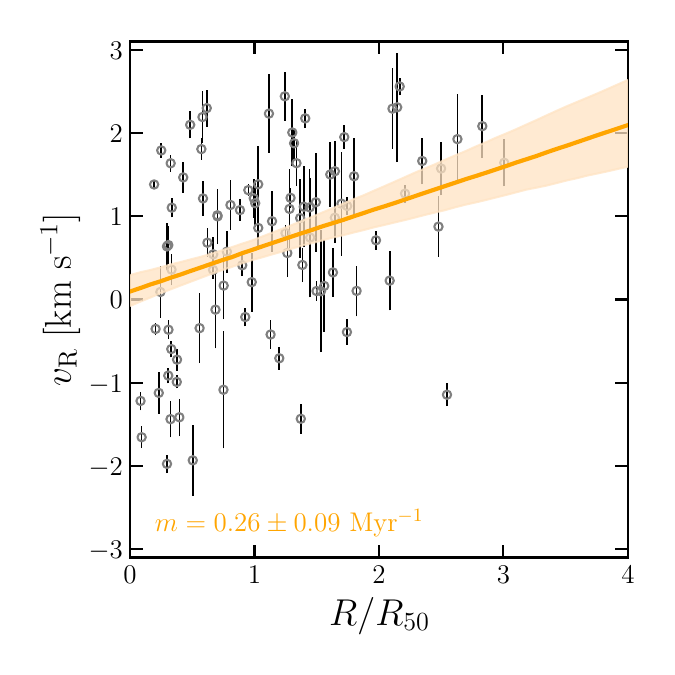} \\ \vspace{-.425cm}
    % NGC 4103
    \includegraphics[width=.325\textwidth]{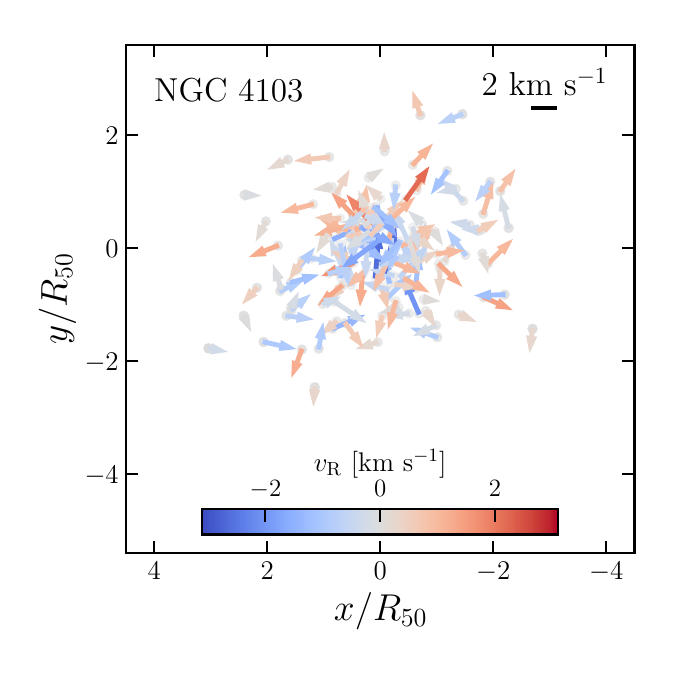}
    \includegraphics[width=.325\textwidth]{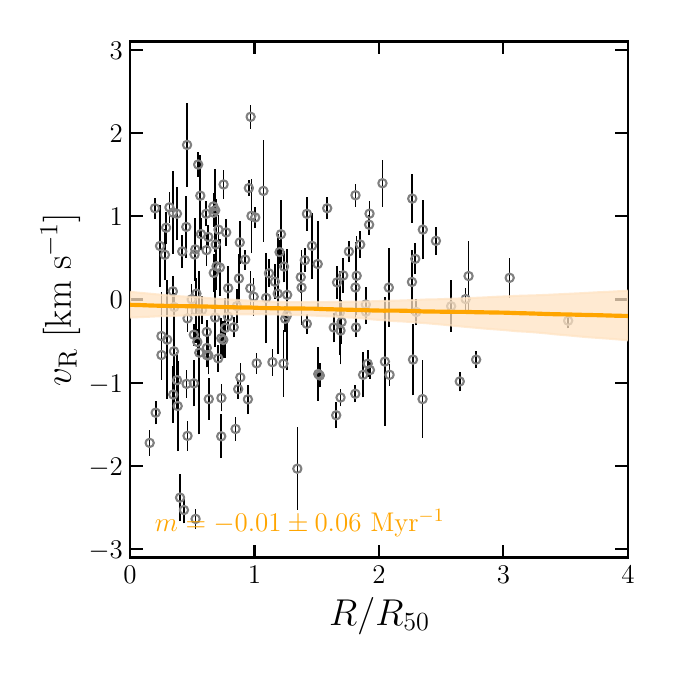} \\ \vspace{-.425cm}
    % LP 2219
    \includegraphics[width=.325\textwidth]{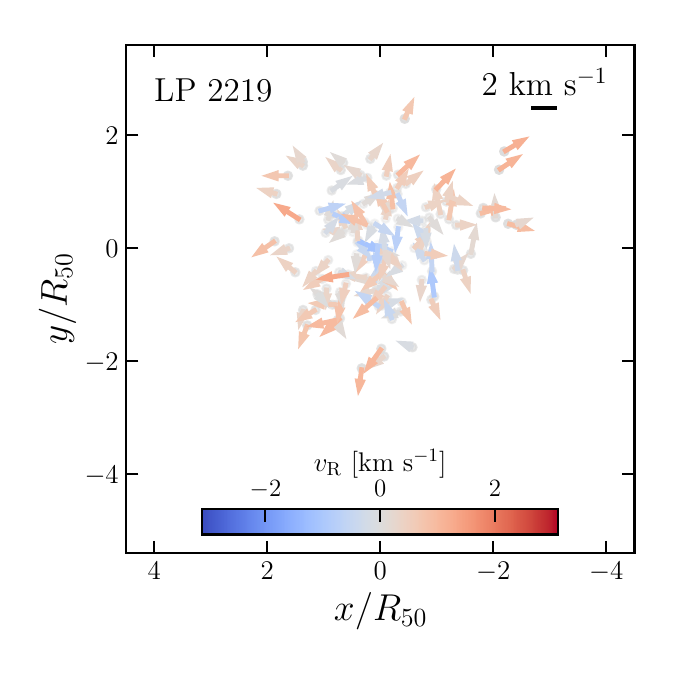}
    \includegraphics[width=.325\textwidth]{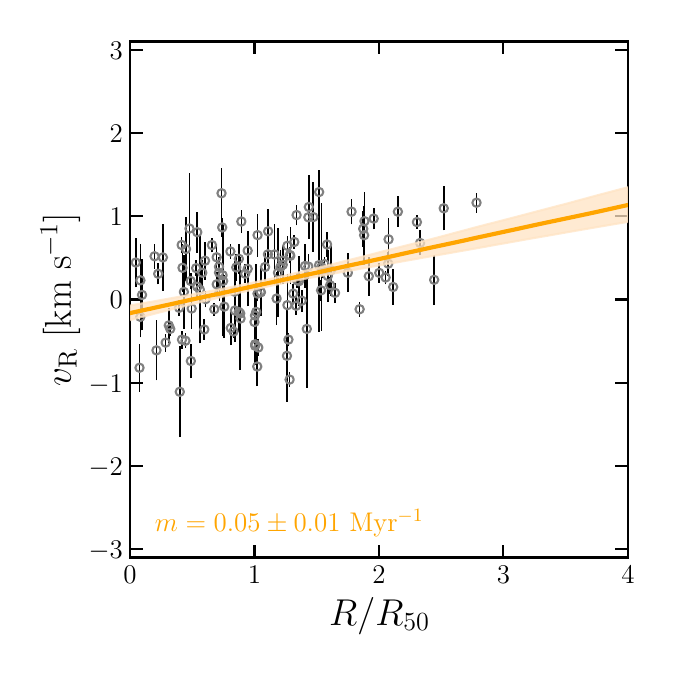} \\ \vspace{-.425cm}
    % NGC 3114
    \includegraphics[width=.325\textwidth]{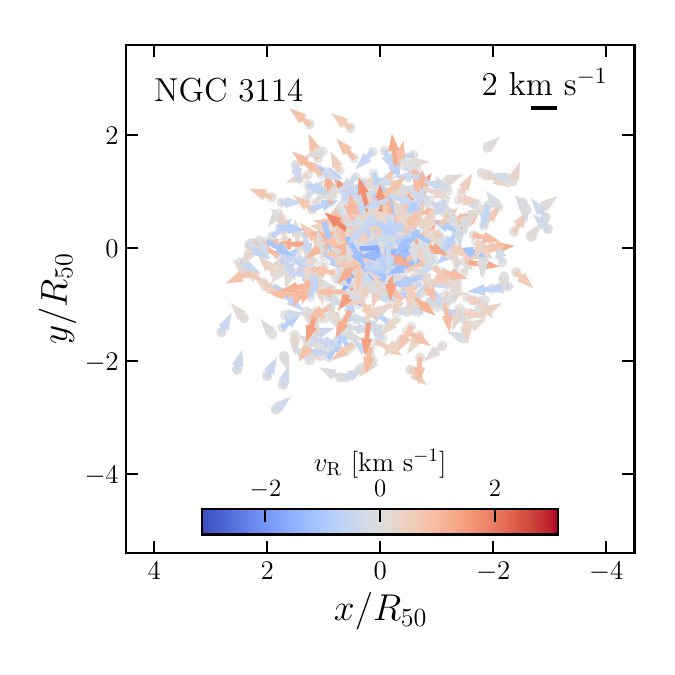}
    \includegraphics[width=.325\textwidth]{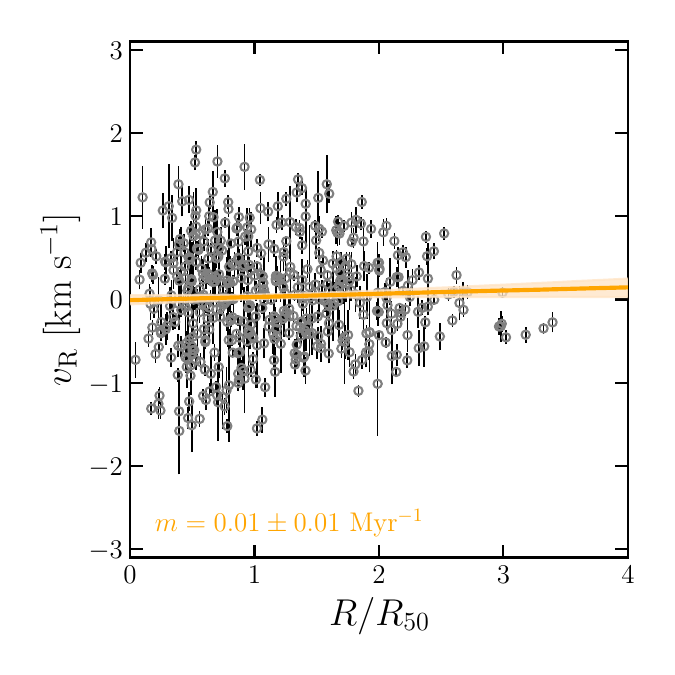}
    \caption{Kinematic properties of NGC~6139, NGC~4103, LP~2219, and NGC~3114 (from top to bottom). The left panels show the spatial distribution of members in Cartesian coordinates normalized to the radius enclosing half of the members, with arrows showing the velocity vectors on the plane of the sky. 
    The arrow lengths are proportional to the speed on the plane of the sky (velocity scale reported in the top-right corners), while their colors map the radial component ($v_{\rm R}$) of the velocity. Positive values point outward.
    The right panels show the distribution of members in the $v_{\rm R} - R$ plane. The linear regressions of velocities as a function of the cluster-centric distances are shown (orange lines), as well as the posterior values on the slopes at the bottom.
    }
    \label{fig:streamplot_ASCC113}
\end{figure*}

Fig.~\ref{fig:expansion} shows the ratio between the mean radial velocity
and the radial velocity dispersion (hereafter $\langle v_{\rm R} \rangle / \sigma_{\rm R}$~) as a function of clusters' ages from \citet{cantat-gaudin+20}. This quantity 
provides an indication of the amplitude of the ordered to the disordered motion of stars along the radial component, 
thus directly tracing ongoing expansion or contraction.
A positive value of $\langle v_{\rm R} \rangle / \sigma_{\rm R}$ implies expansion.
% As a prototype of a mildly expanding system ($\langle v_{\rm R} \rangle / \sigma_{\rm R}>0.5$), we show the velocity field of ASCC~113 members (Fig.~\ref{fig:streamplot_ASCC113}).
% A significant expansion pattern can be observed in this cluster along with a clear positive trend between $v_{\rm R}$ and the cluster-centric distance (right-panel of Fig.~\ref{fig:streamplot_ASCC113}), which results to be dominated by stars in the outer regions  (about $>5$ pc).

The first key result highlighted by Fig.~\ref{fig:expansion} is that 
about $80\%$ of clusters younger than $\sim30$ Myr show positive $\langle v_{\rm R} \rangle / \sigma_{\rm R}$ values. 
More in general, the total fraction of young ($<30$~Myr) systems having positive $\langle v_{\rm R} \rangle / \sigma_{\rm R}$ at the $3\sigma$ level is 43\% (see Table~\ref{tab:expanding_clusters_diffAges} for the fraction of expanding systems in different age bins).
The distribution attains maximum values $\langle v_{\rm R} \rangle / \sigma_{\rm R}=1.5-2$ (such clusters have typically $\langle v_{\rm R}\rangle\sim2$ km s$^{-1}$) for clusters with age $\sim10$ Myr. Then it progressively decreases. 
For clusters older than $\sim 30$ Myr, the distribution of $\langle v_{\rm R} \rangle / \sigma_{\rm R}$ flattens around 0 and it shows an intrinsic standard deviation of about 0.2 (corresponding to $\langle v_{\rm R}\rangle$ $\lesssim0.1$ km s$^{-1}$). 
Such a clear trend allows us to identify for the first time the timescale (of about 30 Myr) during which expansion plays a significant role in the overall cluster kinematics.
We verified that these results do not change significantly if 
different age estimates were adopted
(see Section~\ref{appendix:different_age_estimate}).

In the left panel of Fig.~\ref{fig:expansion} clusters are color-coded according to the number of member stars.
We note that the amplitude of the scatter around zero observed in systems older than 30~Myr strongly depends on the number of members identified.
Results of numerical experiments of equilibrium stellar clusters (see Section~\ref{appendix:montecarlo}) suggest that the observed spread around zero can be largely explained in terms of statistical fluctuations due to a low number of stars (grey shaded area in Fig.~\ref{fig:expansion}).
This in turn confirms that the distribution observed for older clusters is consistent with what is expected for systems in equilibrium.

Most of the young ($<30$ Myr) clusters with clear ongoing expansion are characterized by some of the largest values (up to $\sim10$ pc) of $R_{\rm 50}$ (Fig.~\ref{fig:expansion}, middle panel). We note that among the expanding 
systems, those with smaller $R_{\rm 50}$ values preferentially attain smaller
$\langle v_{\rm R} \rangle / \sigma_{\rm R}$ by up to a factor of two. Finally, old ($>30$ Myr) expanding clusters have mostly fewer members and larger extensions suggesting that they might have been expanding for several tens of Myr. 

Furthermore, we performed a linear fit to the distribution of $v_{\rm R}$ as a function of the cluster-centric distance for each cluster and we derived the angular coefficient $m$ and its relative error $\sigma_{\rm m}$.
Interestingly, we found that the majority of the expanding clusters show a positive and significant ($m/\sigma_{\rm m}\geq 3$) ranking in their $v_{\rm R}$ distributions as a function of their positions in the cluster (see the right panel in Fig.~\ref{fig:expansion}). 
These patterns could suggest that young expanding clusters are likely losing a fraction of their original mass as stars in the outskirts become unbound.
However, we note in passing that this does not necessarily imply that all the expanding clusters will become unbound.
In fact, stars in the external regions moving away from the cluster 
can produce a positive slope even if the inner parts are not expanding.

Finally, we present the kinematic properties of a few proto-typical systems in Figure~\ref{fig:streamplot_ASCC113}.
We selected two young ($<30$~Myr) clusters, namely NGC~6193, and NGC~4103, and two older systems, LP~2219, and NGC~3114. 
Within each group, we picked one system showing significant evidence of expansion (NGC~6193, and LP~2219), whereas the other one is compatible with equilibrium (NGC~4103, and NGC~3114).
Relevent cluster properties are reported in Table~\ref{tab:properties_cluster_example} for reference.

\renewcommand{\arraystretch}{1.5}
\begin{table}[!th]
    \centering
    \caption{The main properties of the four clusters whose kinematic features are shown in Figure~\ref{fig:streamplot_ASCC113}. 
    } \label{tab:properties_cluster_example}
    \begin{tabular}{llcrr}
    \hline
    Cluster name & $\langle v_{\rm R} \rangle / \sigma_{\rm R}$ & $N_\star$ & Age [Myr] & $R_{\rm 50}$ [pc] \\
    \hline \hline
    NGC~6193& $0.62^{+0.12}_{-0.13}$ & 93 & 5 & $2.0^{+0.1}_{-0.2}$ \\
    NGC~4103& $-0.04^{+0.08}_{-0.08}$ & 148 & 21 & $1.7^{+0.2}_{-0.1}$ \\
    LP~2219& $0.56^{+0.10}_{-0.11}$ & 128 & 126 & $6.8^{+0.3}_{-0.5}$ \\
    NGC~3114& $0.08^{+0.05}_{-0.05}$ & 501 & 145 & $4.5^{+0.2}_{-0.1}$ \\
    \hline
    \end{tabular}
    \begin{tablenotes}
    \item[]\textbf{Notes.} From left to right, cluster name, the ratio between the mean radial velocity and the radial velocity dispersion, number of members, cluster age \citep[according to][]{cantat-gaudin+20}, and radius enclosing half of the members. Errors on $\langle v_{\rm R} \rangle / \sigma_{\rm R}$ are directly obtained from the MCMC sampling of the posterior distribution, whereas errors on $R_{\rm 50}$ are obtained by bootstrap resampling the radial distribution of stellar members. The values reported correspond to the 16th and 84th percentiles of the distributions (i.e. corresponding to the $1\sigma$ value if the distributions were Gaussian).
    \end{tablenotes}
\end{table}

In particular, the left panels of Fig.~\ref{fig:streamplot_ASCC113} show the spatial distribution of members used to compute $\langle v_{\rm R} \rangle / \sigma_{\rm R}$ (see Fig.~\ref{fig:expansion}), as well as the velocity vectors on the plane of the sky. Arrow lengths are proportional to their total speeds whereas the color coding traces the amplitude of the radial velocity component (Fig.~\ref{fig:streamplot_ASCC113}). Right panels, on the other hand, show the distribution of individual velocities as a function of the cluster-centric distance. The linear regression used to obtain the $m/\sigma_{\rm m}$ parameter (see Fig.~\ref{fig:expansion}) is also shown and the value of the slope is reported.
To ease the comparison, spatial coordinates are scaled to $R_{\rm 50}$ for each cluster, and their velocities are shown in the same velocity scale.

Expanding clusters (NGC~6193, and LP~2219) show a preferential alignment of velocity vectors along the positive (i.e. pointing outward) radial direction, as highlighted by the color coding (Fig.~\ref{fig:streamplot_ASCC113}). This feature is particularly evident in the external regions. A clear trend is also observed in $v_{\rm R}$ vs $R$ distribution, and it is reflected in the positive value of the slope from the linear regression (as already pointed out in Fig.~\ref{fig:expansion}).
% In addition, NGC~6139 shows a higher expansion velocity compared to LP~2219, as expected since it is younger (see Fig.~\ref{fig:expansion} and the cluster properties listed in Table \ref{tab:properties_cluster_example}).

Non-expanding clusters (NGC~4103, and NGC~3114), on the other hand, present radial velocities that are scattered around zero, without any preferential alignment along the radial direction or significant radial trend. All these features are consistent with the systems being in equilibrium.

\begin{figure}[!t]
    \centering
    \includegraphics[width=0.4\textwidth]{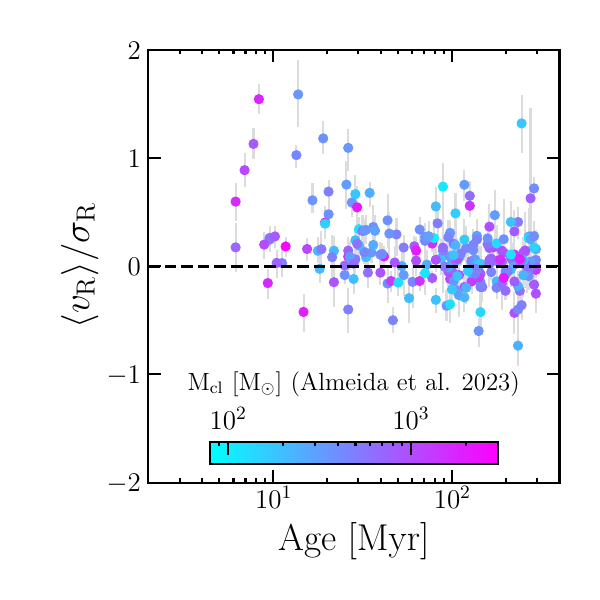}
    \caption{The distribution of $\langle v_{\rm R} \rangle / \sigma_{\rm R}$ as a function of the age for the 227 clusters in common with \citet{almeida_etal2023}. The color coding depicts the cluster mass M$_{\rm cl}$.}
    \label{fig:almeida23_mass}
\end{figure}

Lastly, we looked for any dependence of the expansion properties on the cluster masses. In particular, we crossmatched our cluster catalog with the recent compilation of cluster masses provided by \citet{almeida_etal2023}.
We found 227 clusters in common. For those clusters, we show $\langle v_{\rm R} \rangle / \sigma_{\rm R}$  as a function of cluster age, color-coded according to the mass reported by \citet[][see Fig.~\ref{fig:almeida23_mass}]{almeida_etal2023}.
Qualitatively, expansion affects clusters of any mass, from about $10^2$~M$_\odot$ to a few $10^3$~M$_\odot$. 
To elaborate more on this point, we split the population of clusters younger than 30~Myr (45 clusters) into two sub-samples: expanding (27 clusters, about 60\% of the sample), and non-expanding clusters (18 out of 45 systems, i.e. the 40\%). For the purpose of this analysis, we classified a cluster to be expanding if the expansion signal is significant at the $1\sigma$ level. 
The only purpose here is to obtain two almost equally populated samples of clusters.
We then compared the mass distribution of the two sub-populations with each other and with the full sample of young clusters. A Kolmogorov-Smirnov test shows that there is no significant statistical difference between the three populations.
Although based on a few dozen clusters, this result suggests that the physical processes driving the expansion of star clusters in their early stages of formation and evolution are effective irrespective of the cluster mass.
\FloatBarrier

\section{Comparison with numerical simulations}\label{sec:numerical_simulations}
\subsection{$N$-body simulations of cluster formation}
\begin{figure}[!ht]
    \centering
    \includegraphics[width=0.4\textwidth]{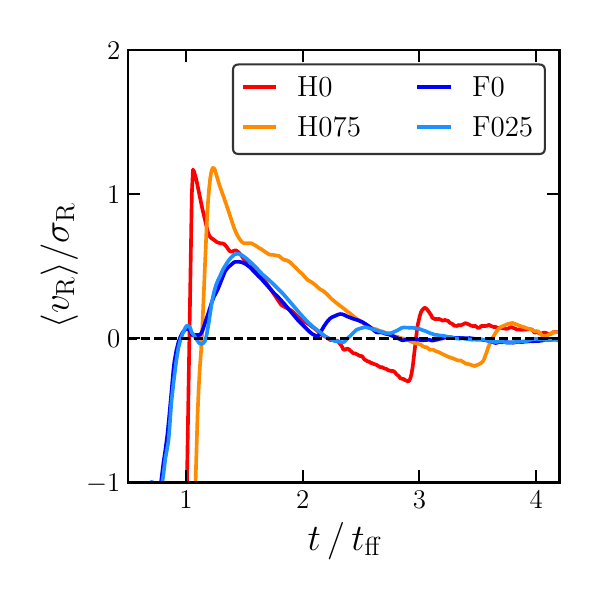}
    \caption{Time evolution of the ratio between the mean radial velocity and the radial velocity dispersions for a selection of models from \cite{livernois2021} for all stars within the tidal radius.}
    \label{fig:theo_comp}
\end{figure}

In this section, we present 
a brief analysis of a few $N$-body simulations of young star clusters undergoing the violent relaxation phase and evolving toward their final virial equilibrium state. The goal here is to illustrate the time evolution of their global expansion pattern and establish a general connection with the observational results presented in the previous sections. The simulations considered here are part of 
the suite discussed in detail in \citet{livernois2021} who explored the early evolution of systems starting with the homogeneous or fractal spatial distribution and investigated the role of initial rotation on the cluster's early evolutionary phases. 
Here we focus on four models: two models with homogeneous initial spatial distributions, one without and one with rotation 
\citep[hereafter referred to as H0 and H075 respectively; see ][for further details about the initial conditions]{livernois2021}
and two models with an initial fractal spatial distribution without and with the initial rotation 
(hereafter F0 and F025).
Fig.~\ref{fig:theo_comp} shows the time evolution of $\langle v_{\rm R} \rangle / \sigma_{\rm R}$ for the selected models as obtained by using all stars in the system within the tidal radius. 
All models show an initial contracting phase followed by significant expansion. 
The initially homogeneous models display more rapid and extreme collapse and expansion phases than the initially fractal models as the clumps within the latter models merge and interact with other clumps before arriving at the center of the system.
Although these specific models do not reach the extreme values of $\langle v_{\rm R} \rangle / \sigma_{\rm R}$ found in our study (Fig.~\ref{fig:expansion}), the range of expansion values found in these simulations spans those attained by most of the observed clusters, thus suggesting that violent relaxation can play a key role in triggering and driving early cluster expansion. 
We emphasize that the idealized models presented here  are included just to illustrate the general kinematic behavior during these early evolutionary phases and they are not meant to provide a detailed fit to the observational data. Different initial conditions and more realistic simulations including additional processes such as gas expulsion and mass loss due to stellar evolution might be necessary to reach the most extreme values found in the observational sample and to constrain the range of timescales of the early expansion, of the settling to the final equilibrium as well as the timescale associated with the possible cluster's dissolution.
We note also that the theoretical lines shown in Fig.~\ref{fig:theo_comp} represent the evolutionary path of the $\langle v_{\rm R} \rangle / \sigma_{\rm R}$ ratio of clusters surviving the early evolutionary phases.  We point out that it is likely that not all the clusters with age~$<30$~Myr in the observational sample will follow this path as some of them will continue expanding and will eventually dissolve. 

\subsection{Distribution of $\langle v_{\rm R} \rangle / \sigma_{\rm R}$ for star clusters in equilibrium}\label{appendix:montecarlo}

Results presented in Fig.~\ref{fig:expansion} suggest that for $t>30$ Myr the distribution of the $\langle v_{\rm R} \rangle / \sigma_{\rm R}$ is compatible with what is expected for clusters in equilibrium and that the broadening of the distribution might be mostly driven by statistical fluctuations.

To check whether this is indeed the case, we have created 100 random realizations of a population of clusters. Each population contains the same number of clusters as in our observed sample and in each realization the clusters have the same number of stars as in the observed sample. Positions and velocity of stars in each cluster follow those of a King model with central dimensionless potential $W_0 = 5$ (but the results do not have any significant dependence on the particular model adopted).
For each realization, we have then calculated the dispersion in the distribution of the values of $\langle v_{\rm R} \rangle / \sigma_{\rm R}$ of the clusters in the sample.
The average value of the dispersion found in the 100 realizations is equal to about 0.11 and is shown as a grey-shaded area in Fig.~\ref{fig:expansion}. The comparison demonstrates that 1) the spread observed for older systems can be largely accounted for by statistical fluctuations and 2) equilibrium models cannot account for large positive $\langle v_{\rm R} \rangle / \sigma_{\rm R}$ (i.e. expansion) observed in clusters with $t < 30$ Myr and that those systems are therefore out of equilibrium.

We note, however, that the observed spread is larger (about a factor 2) than that derived from the simulations. While these residuals might suggest that some of the clusters older than $\sim30$ Myr are still oscillating around equilibrium configurations, it is important to emphasize that in the analysis of the $\langle v_{\rm R} \rangle / \sigma_{\rm R}$ from the numerical realizations of cluster populations, not all the possible effects that can determine the spread of this quantity are included. They do not account, for example, for a number of possible uncertainties associated with the observational data such as the uncertainties induced by wrong LOS systemic velocities (mainly due to low-number statistics). Further investigation is needed to properly study this issue.

\section{Testing the impact of different age estimates on the expansion pattern}\label{appendix:different_age_estimate}
The reference age compilation adopted in this analysis is the one by \citet{cantat-gaudin+20}.
Here we test the robustness of our results against different compilations of clusters ages from literature. In particular, we used cluster ages obtained by
\citet{Kharchenko_etal13}, \citet{bossini_etal2019}, and \citet{dias_etal2021} who determined ages for several star clusters in our Galaxy through isochrone fitting. We note that \citet{Kharchenko_etal13} exploited pre-main-sequence stars as a further age indicator to obtain more reliable ages in the young end. \citet{hunt&reffert2023}, performed an all-sky search for open clusters and they provided ages for every cluster in their sample obtained by means of a convolutional neural network.
We crossmatched the sample of 509 clusters adopted in this study with these catalogs.

\begin{figure*}[!th]
    \centering
    \includegraphics[width=.34\textwidth]{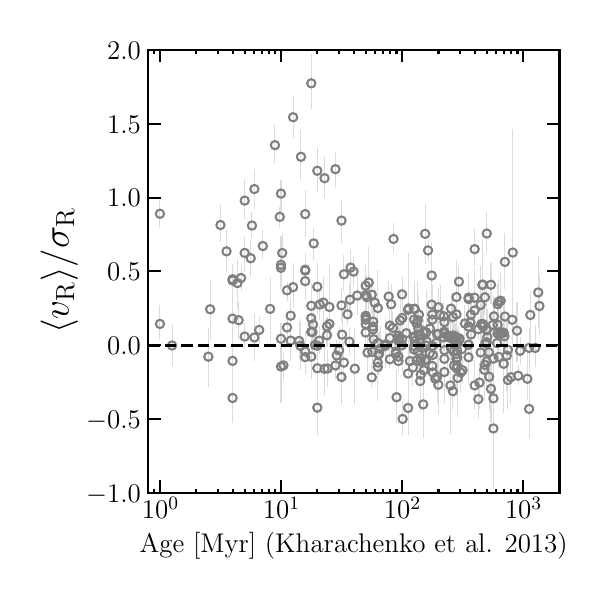}
    \includegraphics[width=.34\textwidth]{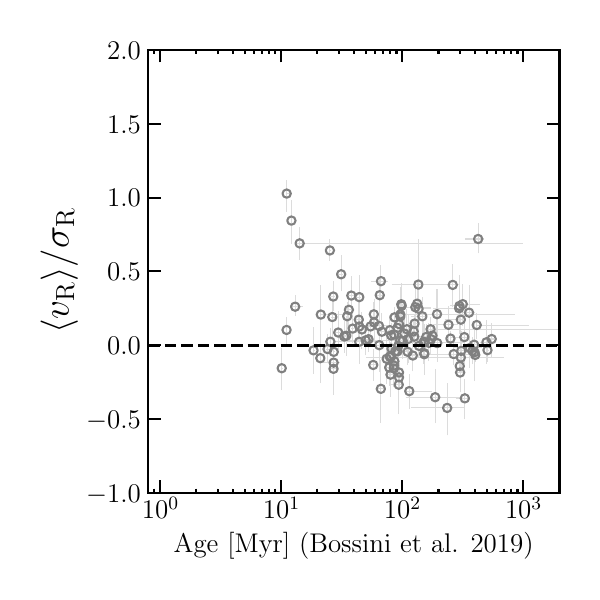}
    \includegraphics[width=.34\textwidth]{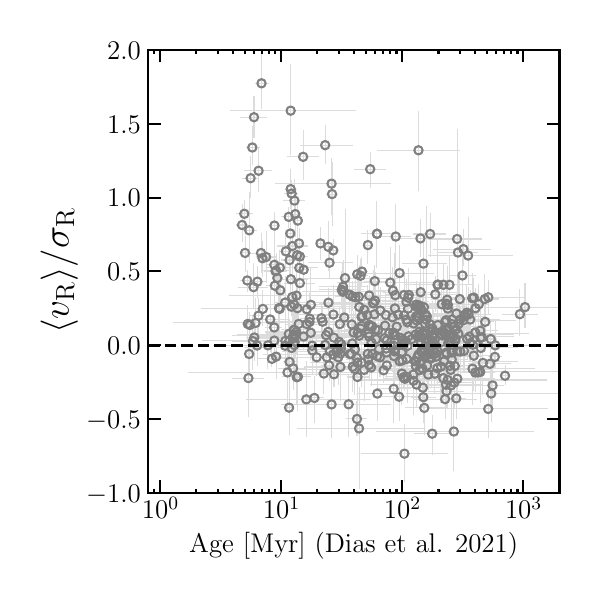}
    \includegraphics[width=.34\textwidth]{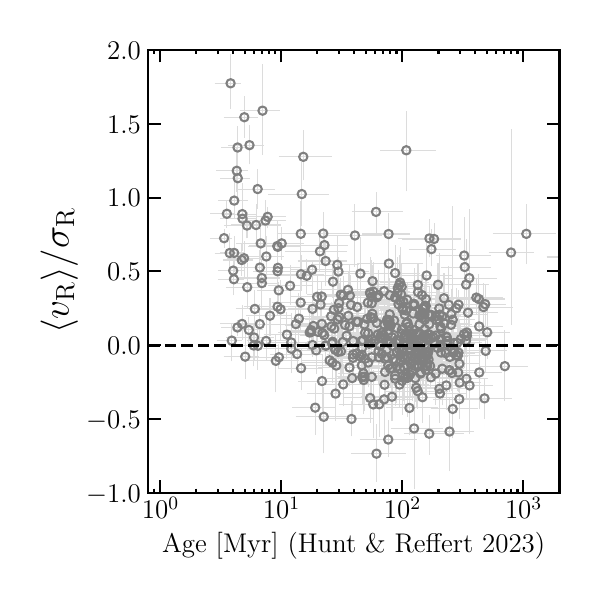}
    \caption{Mean radial velocity to radial velocity dispersion ratio as a function of cluster age for different age estimates: from left to right \citet{Kharchenko_etal13}, \citet{bossini_etal2019}, \citet{dias_etal2021}, and \citet{hunt&reffert2023}. 
    Errors on the $y$-axis are obtained from the MCMC sampling of the posterior, while those on the $x$-axis (if present) are age errors provided by the different catalogs.
    The zero-level expansion is marked by the dark, dashed line.}
    \label{fig:comparison_ages_HR23}
\end{figure*}

Figure~\ref{fig:comparison_ages_HR23} shows the distribution of $\langle v_{\rm R} \rangle / \sigma_{\rm R}$ as a function of age (as in Figure~\ref{fig:expansion}) for the clusters in common with the four compilations.
Also, in Table~\ref{tab:expanding_clusters_diffAges} we report the fractions (as well as the total numbers) of clusters exhibiting evidence (at the $3\sigma$ level) for expansion in different age bins.
Although different catalogs span different age ranges for the same clusters, the trend of expanding clusters for ages below $\simeq30$ Myr is clearly visible in all the catalogs.
This test demonstrates that the results presented in the manuscript are robust against different age estimates, also in terms of time scale during which expansion has an important role in cluster kinematics.

\begin{table*}[!th]
    \centering
    \caption{The fraction of expanding clusters} \label{tab:expanding_clusters_diffAges}
    \begin{tabular}{llcrr}
        \hline
        Ref. ages compilation & $\leq$ 10 Myr& $(10;30]$ Myr & $(30;50]$ Myr & $>50$ Myr \\
        \hline \hline
        \cite{Kharchenko_etal13}        & 17/32 ($\simeq$53\%) & 13/44 ($\simeq$30\%) & 4/14 ($\simeq$29\%) & 9/219 ($\simeq$4\%) \\
        \cite{bossini_etal2019}         & $-$ & 6/18 ($\simeq$33\%) & 1/12 ($\simeq$8\%) & 2/93 ($\simeq$2\%)  \\
        \cite{cantat-gaudin+20}         & 15/28 ($\simeq$54\%) & 43/108 ($\simeq$40\%) & 5/53 ($\simeq$9\%) & 14/320 ($\simeq$4\%)  \\
        \cite{dias_etal2021}         & 21/41 ($\simeq$51\%) & 28/83 ($\simeq$34\%) & 4/43 ($\simeq$9\%) & 12/226 ($\simeq$5\%)  \\
        \text{\cite{hunt&reffert2023}}  & 34/51 ($\simeq$67\%) & 15/57 ($\simeq$26\%) & 1/39 ($\simeq$3\%) & 11/302 ($\simeq$4\%)  \\
        \hline
    \end{tabular}
    \begin{tablenotes}
    \item[]\textbf{Notes.} The fraction and the absolute number of clusters that show significant expansion (at the $3\sigma$ level). Values are reported in four age bins, namely $\leq10$ Myr, between $10-30$ Myr, between $30-50$ Myr, and $>50$ Myr, according to five different age compilations (left column). The only exception is the catalog of \citet{bossini_etal2019} for which we did not have any cluster younger than 10 Myr in common.
    \end{tablenotes}
\end{table*}

\section{Comparison with previous works}\label{sec:comparison_previous_works}
\citet{kuhn_etal2019} investigated the expansion properties of twenty-eight stellar clusters and associations, concluding that at least 75\% of clusters in their sample are expanding. 
Among the clusters in common, we find a significant expansion for NGC~1893, NGC~2244, and NGC~2362 and a mild expansion for IC~348, and NGC~6231, consistent with the results by \citet{kuhn_etal2019} for these systems. 

Recently, \citet{Guilherme-Garcia_etal2023} investigated the internal kinematics of many open clusters. In particular, they identified expansion in 14 clusters with 15 more candidates. 
Among the 14 expanding clusters, 5 were not included in our study, namely Alessi~13, Aveni Hunter~1, Collinder~132 (as they have less than 30 members in our catalog, see Section~\ref{sec:membership}), Ruprecht~98, and Stock~1 (as they are older than 300 Myr according to \citealt{cantat-gaudin+20}). For all the others, we confirm their state of significant expansion, with the only exception of BH~164 for which we derived $\langle v_{\rm R} \rangle / \sigma_{\rm R} \simeq +0.13^{+0.10}_{-0.11}$.
Among the 15 candidates, according to our analysis 
we confirm that IC~1805 ($\langle v_{\rm R} \rangle / \sigma_{\rm R} \simeq +1.77^{+0.20}_{-0.17}$), Roslund~2 ($\langle v_{\rm R} \rangle / \sigma_{\rm R} \simeq +1.06^{+0.14}_{-0.14}$), Trumpler~16 ($\langle v_{\rm R} \rangle / \sigma_{\rm R} \simeq +0.81^{+0.15}_{-0.10}$), and vdBergh~92 ($\langle v_{\rm R} \rangle / \sigma_{\rm R} \simeq +1.13^{+0.15}_{-0.14}$) are expanding, whereas ASCC~127 ($\langle v_{\rm R} \rangle / \sigma_{\rm R} \simeq -0.02^{+0.15}_{-0.13}$), and BH~99 ($\langle v_{\rm R} \rangle / \sigma_{\rm R} \simeq -0.075^{+0.081}_{-0.080}$) are not.
In addition to differences in specific systems, we note that in our analysis we found a significantly larger sample of expanding clusters than \citet{Guilherme-Garcia_etal2023}. This is probably due to different approaches and membership probabilities that are based on Gaia DR3 data (Section~\ref{sec:membership}) in our case and DR2 in the case of \citet{Guilherme-Garcia_etal2023}.

Besides dedicated studies \citep[such as][]{kuhn_etal2019,Guilherme-Garcia_etal2023}, evidence of expansion was found in several works.
For instance, \citet{bravi_etal2018} investigated the kinematical properties of four young ($\gtrsim30$ Myr) open clusters, namely IC~2602, IC~2391, IC~4665, and NGC~2547. They found that all the clusters but IC~4665 (for which they only put an upper limit) are super virial, concluding that this is consistent with the residual gas expulsion scenario. Clusters indeed expand eventually returning to an equilibrium state after unbound stars are dispersed. This process might take several tens of system crossing times \citep{baumgardt_kroupa2007}.
Consistently, we found that all these four clusters show slow expansion speeds (typically $\langle v_{\rm R} \rangle / \sigma_{\rm R} \lesssim 0.3 \pm 0.1$), suggesting they might be close to equilibrium.

\citet{lim_hong_etal2020} studied the star-forming region W4, where the cluster IC~1805 is located. They found that the cluster is composed of an isotropic core and an external region showing clear evidence of expansion. 
These features suggest that the cluster is experiencing expansion after an early, initial collapsing phase \citep{lim_hong_etal2020}.
We also found a strong indication of expansion in IC~1805, although we did not find evidence for a central isotropic core in the cluster.

The cluster NGC~2244 lies at the center of the Rosette Nebula. The region shows a complex interplay between stellar and gas kinematics, and feedback-driven star formation in substructured environments. Both rotation and expansion were found in NGC~2244 \citep{lim_naze_hong_etal2021}.
In our study we also found NGC~2244 to be significantly expanding ($\langle v_{\rm R} \rangle / \sigma_{\rm R} \simeq +0.98^{+0.14}_{-0.13}$).

\citet{pang2021} investigated the connection between the cluster's internal kinematics and their morphology. They found younger clusters to exhibit filament-like substructures while older ones show tidal-tail features. The majority of the systems are inferred gravitationally unbound and expanding \citep{pang2021}. 
In the present study, mild expansion has been directly detected in NGC~2422, NGC~2451a, NGC~2451b, NGC~2232. On the other hand, neither expansion nor contraction has been found in NGC~2516, suggested to be in a super virial state \citep{pang2021}.
Investigating internal kinematic patterns, or dependencies on cluster morphology for each cluster is beyond the scope of this work. We defer this topic to a follow-up study.

In summary, the excellent agreement with previous, detailed studies that focussed on a handful of systems suggests that our results are solid and that we are effectively probing the expansion of young stellar clusters.

\section{Discussion and Conclusion}\label{sec:conclusions}
We performed a comprehensive analysis of the internal kinematics of young star clusters ($t<300$ Myr) in the Milky Way with the aim of reconstructing the key properties and possible physical mechanisms 
shaping the early cluster expansion. 
We emphasize that this analysis is based on a sample 20 times larger and spanning a cluster age range 60 times larger than previous analyses \citep[see e.g.][]{kuhn_etal2019},
thus enabling for the first time the possibility of constraining the timescale during which expansion has a dominant impact on cluster kinematics and the fraction of stellar systems significantly affected by the expansion.

Our analysis reveals a clear trend in which the fraction of expanding clusters increases for younger clusters (see Table~\ref{tab:expanding_clusters_diffAges}): a significant fraction of clusters younger than $\sim$30 Myr are characterized by a significant expansion attaining values as large as $\langle v_{\rm R} \rangle / \sigma_{\rm R}=1.5-2$. On the contrary, older clusters (t > 30 Myr) are mostly consistent with what is expected for systems in equilibrium.
% We found that up to $80\%$ (median value) of clusters younger than $\sim30$ Myr are characterized by a significant expansion attaining values as large as $\langle v_{\rm R} \rangle / \sigma_{\rm R}=1.5-2$~. On the contrary, older clusters ($t>30$ Myr) are mostly consistent with what is expected for systems in equilibrium.
While it would be tempting to interpret the $\langle v_{\rm R} \rangle / \sigma_{\rm R}$ distribution as a function of time as an evolutionary sequence, we stress here that 
it represents an instantaneous picture of the current properties of stellar clusters and not a time evolution pattern.
The results presented in this work do not significantly change if different catalogs of cluster ages were adopted.

A general comparison of the evolution of $\langle v_{\rm R} \rangle / \sigma_{\rm R}$ in $N$-body simulations following the violent relaxation
and early dynamics of star clusters shows values of $\langle v_{\rm R} \rangle / \sigma_{\rm R}$
spanning those found in most of the observed clusters. More realistic simulations including additional processes (such as gas expulsion and mass loss due to stellar evolution) and exploring a broader range of initial conditions will be required to explain the most extreme cases of expanding systems ($\langle v_{\rm R} \rangle / \sigma_{\rm R}>1.5-2$) sampled by our observational analysis and for a more detailed and quantitative comparison of the observed kinematic patterns and the timescales associated with the early evolution.
Finally, we note that extremely young clusters (with ages smaller than a few Myr) not included in our sample would be necessary to probe the kinematic patterns associated with the systems' very early dynamics. The lack of these systems is likely a selection effect, as these very young systems are probably embedded clusters and could hardly be observed with Gaia. In this respect, future data from infra-red surveys would provide relevant insights into the embedded cluster population
\citep[see for example the VISIONS survey,][]{meingast_etal2023_VISIONS} 
and allow us to build a more complete dynamical picture of the evolution of these systems.

\begin{acknowledgements}
We thank the anonymous referee for their insightful comments which improved the quality of the paper.
The data underlying this article will be shared on reasonable request to the corresponding author.
This work uses data from the European Space Agency (ESA) space mission Gaia. Gaia data are being processed by the Gaia Data Processing and Analysis Consortium (DPAC). Funding for the DPAC is provided by national institutions, in particular, the institutions participating in the Gaia Multi-Lateral Agreement (MLA).
A.D.C. and E.D. acknowledge financial support from the project Light-on-Dark granted by MIUR through PRIN2017-2017K7REXT contract. E.D. acknowledges financial support from the Fulbright Visiting Scholar program 2023. A.D.C. and E.D. are also grateful for the warm hospitality of Indiana University where part of this work was performed.
E.V. acknowledges support from the John and A-Lan Reynolds Faculty Research Fund.
\end{acknowledgements}

\bibliographystyle{aa}
\bibliography{bibliography}

% \appendix
% \section{Additional figures not included in the main paper}
% These figures are all relevant to the discussion on the dependence on the cluster mass and are included for \emph{internal} reference/use.
% \begin{figure*}
%     \centering
%     \includegraphics[width=0.30\textwidth]{plots/cumulative_mass_distr_almeida23.pdf}
%     \includegraphics[width=0.32\textwidth]{plots/voutSigma_R95Rj_bins.pdf}
%     \includegraphics[width=0.32\textwidth]{plots/voutSigma_R95Rj.pdf}
%     \caption{my Caption}
% \end{figure*}
\end{document}